\documentclass[twocolumn,preprintnumbers,amsmath,amssymb]{revtex4}

\def\be{\begin{equation}}
\def\ee{\end{equation}}

\def\be{\begin{equation}}
\def\ee{\end{equation}}

\def\be{\begin{equation}}
\def\ee{\end{equation}}

\usepackage[dvipdfmx]{graphicx}
\usepackage{here,braket,booktabs}
\usepackage{comment}
\usepackage{epsf}
\usepackage{amsmath}
\usepackage{graphics}
\usepackage{amsfonts}
\usepackage{amssymb}
\usepackage{latexsym}
\usepackage{color}
\usepackage{natbib}
\usepackage[colorlinks]{hyperref}

\input{colordvi.tex}
\usepackage{feynmp}

\begin{document}
\preprint{RESCEU-10/18}

\title{Probing microstructure of black hole spacetimes with gravitational wave echoes}

\author{Naritaka Oshita}
\affiliation{
  Research Center for the Early Universe (RESCEU), Graduate School
  of Science,\\ The University of Tokyo, Tokyo 113-0033, Japan and\\
   Department of Physics, Graduate School of Science, The University of Tokyo, Tokyo 113-0033, Japan}

\author{Niayesh Afshordi}
\affiliation{
 Department of Physics and Astronomy, University of Waterloo, Waterloo, ON, N2L 3G1, Canada and\\ Perimeter Institute for Theoretical Physics, 31 Caroline St. N., Waterloo, ON, N2L 2Y5, Canada 
}



\begin{abstract}

Quantum nature of black hole horizons has been a subject of recent interest and scrutiny. In particular, a near-horizon quantum violation of the equivalence principle has been proposed as a resolution of the black hole information paradox. Such a violation may lead to a modified dispersion relation at high energies, which could become relevant due to the intense gravitational blueshift experienced by ingoing gravitational waves.
We investigate the ringdown for a perturbed black hole with such a modified dispersion relation
and find that infalling gravitational waves are partially reflected near the horizon.
This results in the appearance of late-time {\it echoes} in the ringdown phase of black hole merger events, with similar properties to those (arguably) seen in the Advanced LIGO observations. Current measurements suggest a Lorentz-violation scale of $10^{13 \pm 2}$ GeV for gravitational waves, with comparable dissipation and dispersion. 
Therefore, if confirmed, black hole ringdown echoes probe the microstructure of horizons and thus can test Lorentz-violating UV completions.
\end{abstract}


\maketitle

\section{Introduction}
Reconciliation of Einstein's theory of general relativity with quantum mechanics
is one of the deepest mysteries in theoretical physics, having led to a multitude of proposals for a theory of
quantum gravity. A quantum theory of gravity is expected to provide a microscopic picture of spacetime. One litmus test is an explanation of 
the laws of black hole thermodynamics, which suggest a microscopic entropy associated with event horizons \cite{Bekenstein:1972tm,Bekenstein:1973ur,Bekenstein:1974ax}:
the number of black hole microstates should be $N_{\text{BH}} \sim e^{\mathcal{A}/4G}$,
where $\mathcal{A}$ is the area of black hole horizon and $G \equiv \ell_{\text{Pl}}^2$ is the
Planck area that is around $10^{-70} \ \text{m}^2$. Therefore, black hole entropy (Bekenstein-Hawking entropy)
is given by $S_{\text{BH}} \equiv \ln{N_{\text{BH}}} \sim \mathcal{A}/4G$, which can be derived
from a semiclassical analysis of the gravitational path integral \cite{Gibbons:1976ue}.
However, the nature of these black hole horizon microstates has remained illusive,
an issue of which the resolution will shed light onto the darkest core of quantum gravity.

It has been recently proposed that ringdown gravitational waves (GWs) may be useful to test the structure of black hole horizons \cite{Vicente:2018mxl,Bueno:2017hyj,Maselli:2017cmm,Cardoso:2016oxy,Cardoso:2016rao,Wang:2018gin}.
This structure may arise in exotic compact objects
such as wormholes \cite{Visser:1995cc}, gravastars \cite{Mazur:2001fv}, fuzzballs \cite{Mathur:2005zp}, 
firewalls \cite{Almheiri:2012rt}, 2-2 holes \cite{Holdom:2016nek}, orbifold membranes \cite{Saravani:2012is}, and a \textit{self-consistent} model of semiclassical black holes \cite{Kawai:2015uya}.
In this paper, we study the possibility that  microstructure on black hole horizons can lead to a
modification of the dispersion relation (DR) of GWs, which can be probed
with the observation of ringdown GWs from merger events that lead to the formation of black holes.

Theories of quantum gravity generically modify the microstructure of spacetime on small scales comparable to Planck length $ \ell_{\text{Pl}}$, which may lead to a modification of DR, or broken Lorentz invariance, as we approach Planck energy (e.g., \cite{Padmanabhan:1998jp,Padmanabhan:1998vr}).
The broken Lorentz invariance requires the existence of a preferred coordinate system (or spacetime foliation). For the purpose of this study, we shall assume that this preferred coordinate system is defined by the black hole spacetime Killing vectors, in terms of which the metric is static (or stationary) \footnote{There is no local process that could set up this preferred foliation. However, the nonperturbative quantum gravity effects that lead to the formation of the exotic compact object may also set up the preferential static foliation. Note that a ``static metric" means that the {\it background} metric is static, but of course small perturbations around the static background (ringdown GWs) are dynamical.}.
For simplicity, we shall focus on a nonspinning black hole, and calculate the ringdown GWs in the Schwarzschild coordinate:
$ds^2 = - F(r) dt^2 +F^{-1}(r) dr^2 + r^2 (d\theta^2 + \sin^2{\theta} d \varphi^2),$
where $F (r) \equiv 1-r_{\text{g}}/r$ and $r_{\text{g}}$ is the Schwarzschild radius.
This static solution is, for instance, realized in the Ho\v{r}ava-Lifshitz (HL) gravity without quadratic curvature terms \cite{Kiritsis:2009rx}.

The Bekenstein-Hawking entropy can be derived by using the (Euclidean) Schwarzschild coordinates that possess the conical singularity on its horizon
\cite{Gibbons:1976ue,Gregory:2013hja,Appels:2017xoe}, and the derivation of the Wald entropy \cite{Wald:1993nt}
is based on a Noether charge which is defined on a Cauchy surface not covering the interior region
but only covering the exterior region.
In this sense, the concept of the Bekenstein-Hawking entropy
can be associated with the Schwarzschild coordinate (rather than e.g., Kruskal coordinates,
which is seen by an infalling observer going across the horizon). As such, one may argue that the static Schwarzschild coordinates are the natural choice of coordinates for black hole microstates, which would break Lorentz symmetry. 
In addition, arguments based on the black hole information paradox 
suggest that the black hole interior may not exist (e.g., \cite{Mathur:2009hf,Braunstein:2009my,Almheiri:2012rt}).
According to this hypothesis, the ``no drama'' picture in which an observer
freely falls across the black hole horizon is no longer valid, signaling a quantum violation of Einstein's equivalence principle.
In contrast, if one chose e.g., the Kruskal coordinates as a preferred frame, nothing special would happen in the ringdown GWs from a black hole since there would be no significant blueshift at the horizon.

If a modified DR turns on as frequencies reach Planck energy $E_{\text{Pl}}$, the adiabatic approximation near the horizon (where frequencies diverge) is violated, and thus we can convert ingoing to outgoing modes. 
This leads to a drastic change in the quasinormal modes (QNMs) of black holes,
and in particular, {\it echoes} could show up in the late-time tail of ringdown GWs.
In this manuscript, we focus on lowest-order Planck-suppressed Lorentz-violating theories by studying a modified DR, $\Omega^2 - K^2 = - C^2 K^4/E_{\text{Pl}}^2 - i \gamma K^2 \Omega / E_{\text{Pl}}$, as a toy model to phenomenologically investigate the effects of Lorentz violation on the ringdown GWs. The case of HL gravity will be discussed elsewhere.
In the case of the HL gravity, the preferred frame is not given by the Schwarzschild coordinates (even in the GR limit of the HL gravity) due to the fall of the khronon field into the Killing horizon \cite{Blas:2011ni,Barausse:2011pu}. The infalling khronon field piles up near the universal horizon just like the Killing foliation piles up near the event horizon. In addition, the universal horizon possesses the microscopic degrees of freedom associated with the black hole's entropy \cite{Cropp:2013sea}. In this sense, our model is just a toy model of the HL gravity.

The organization of our paper is as follows.
In Sec. \ref{sec:model} we assume a modified Regge-Wheeler (RW) equation, leading to a modification of DR,
to provide a toy model to describe a Lorentz violation at the Planck scale, and then
we give physically reasonable boundary conditions to solve the modified RW equation.
In Sec. \ref{sec:QNMs} the QNMs of black holes for the modified DR are shown,
and imposing a certain initial data of GWs, we numerically calculate the time-domain function
of ringdown GWs from a perturbed black hole and find out late-time echoes. Furthermore, we
also numerically investigate the reflection rate of infalling GWs near the horizon.
In Sec. \ref{sec:discussions} we roughly discuss some experimental consequences of our proposal.
Finally, we conclude with some final comments in Sec. \ref{sec:conclusions}.

\section{Model and boundary conditions}
\label{sec:model}
\subsection{Toy model to describe the Lorentz violation around the Planck scale}
The ringdown GWs consist of the QNMs of a black hole, which are calculated by solving the RW equation \cite{Regge:1957td,Zerilli:1970se,Zerilli:1971wd}
\begin{equation}
\left[ \frac{\partial^2}{\partial r^{\ast} {}^{2}} + \omega^2 -V_{\ell,s}(r^{\ast}) \right] \psi_s (r^{\ast}, \omega) = 0,
\label{011001}
\end{equation}
with the appropriate boundary condition
\begin{equation}
\displaystyle \lim_{r^{\ast} \to - \infty} \psi_s \sim e^{-i k r^{\ast}}, \ \displaystyle \lim_{r^{\ast} \to + \infty} \psi_s \sim e^{i k r^{\ast}},
\label{011002}
\end{equation}
where $\ell$ is a spherical-harmonics mode, $s$ is the index of the spherical-harmonic expansion, and $V_{\ell,s}(r^{\ast})$ is the RW potential.
Here we define the Schwarzschild frequency, $\omega$, and tortoise wave number, $k$,
which are the conjugates of the Schwarzschild time, $t$, and tortoise coordinate, $r^{\ast}$, respectively.
Substituting $e^{\pm i k r^{\ast}}$ into the RW equation (\ref{011001}), one finds that it 
reduces to the DR: $-k^2 + \omega^2 = 0$ for $r^{\ast} \to \pm \infty$.
Rewriting this DR with the proper (physical) frequency ($\Omega$) and wave number ($K$),
\begin{equation}
\Omega \equiv \omega/ \sqrt{F (r)} \ \text{and} \ K \equiv k/ \sqrt{F(r)},
\label{DR111}
\end{equation} 
one obtains
\begin{equation}
\Omega^2 - K^2 = 0 \ \text{for} \ r^{\ast} \to \pm \infty.
\label{18081803}
\end{equation}

To demonstrate that a UV Lorentz violation could affect
ringdown GWs propagating from a black hole, for an example,
we will use the following modified RW equation,
\begin{widetext}
\begin{equation}
\left[\frac{C^2}{E_{\text{Pl}}^2} F^{-1}(r) \frac{\partial^4}{\partial r^{\ast} {}^{4}}
-i \frac{\gamma \omega}{E_{\text{Pl}}} F^{-1/2} (r) \frac{\partial^2}{\partial r^{\ast} {}^{2}}
+ \frac{\partial^2}{\partial r^{\ast} {}^{2}} + \omega^2 -V_{\ell,s}(r^{\ast}) \right] \psi_s (r^{\ast}, \omega) = 0,
\label{010201}
\end{equation}
\end{widetext}
where $C$ and $\gamma$ are arbitrary parameters, similar to ``lattice size effects'' and ``viscosity'' in ordinary material.
Note that we simply assume (\ref{010201}) here as a toy model to describe lowest-order Lorentz violation in the IR limit. For instance, in the HL gravity,
we also have to take a $\partial_{r^{\ast}}^6$ term into account, and there might be some nontrivial modifications
in the RW potential.
Here, the modification terms, the first and second terms in (\ref{010201}), are negligible for $r^{\ast} \to \infty$,
and (\ref{010201}) reduces to (\ref{011001}) at a distant region
since those terms are suppressed by factors $k/E_{\text{Pl}} \sim \omega/E_{\text{Pl}} \sim \ell_{\text{Pl}} / r_g$ while $F \simeq 1$.
On the other hand, in the near-horizon limit, the modification terms are dominant since
$F^{-1} \simeq e^{-r^{\ast}/r_g +1} \to \infty$ for $r^{\ast} \to - \infty$.
The modified RW equation (\ref{010201}) reduces to the following modified DR in the near-horizon limit
and at a distant region:
\begin{equation}
\frac{C^2}{E_{\text{Pl}}^2} \frac{k^4}{F} + i \frac{\gamma}{E_{\text{Pl}}} \frac{\omega}{F^{1/2}} k^2
-k^2 + \omega^2 = 0 \ \text{for} \ r^{\ast} \to \pm \infty.
\label{18081801}
\end{equation}
Rewriting (\ref{18081801}) with the proper frequency and wave number (\ref{DR111}),
one finds that (\ref{010201}) gives the following quartic DR:
\begin{equation}
\Omega^2 -K^2 = - C^2 K^4/E_{\text{Pl}}^2 -i \gamma K^2 \Omega/E_{\text{Pl}}.
\label{18020401}
\end{equation}
Note that the wave number at which the group velocity becomes zero, $K_{\text{max}}$, is estimated as
$K_{\text{max}} = E_{\text{Pl}}/\sqrt{\gamma^2/2 + 2 C^2}$ (see Appendix \ref{app:a})
and the dissipation effect is adjustable by changing the value of $\gamma$.
Here, we only include the radial derivative terms, the first and second terms in (\ref{010201}),
as modifications of RW equation since they dominate in the near-horizon limit because of the blueshift factor, $1/F (r)$. All modification terms are suppressed by $(GM)^{-1}/E_{\text{Pl}}$ and are negligible far from the black hole.

\subsection{Boundary conditions}
To compute QNMs, we can impose the outgoing modes at the
distant region, $\displaystyle \lim_{r^{\ast} \to \infty} \psi_s \sim e^{i k r^{\ast}}$.
On the other hand, imposing ingoing modes near the horizon is impossible since $F(r)^{-1} = (1-r_{\text{g}}/r)^{-1}$ in (\ref{010201})
diverges at the horizon and the first term becomes dominant, which, as we see below, gives a nonoscillatory solution near the horizon.
The modified RW equation (\ref{010201}), is a fourth-derivative equation, and therefore, it gives four independent solutions.
Using $F(r) \simeq e^{r^{\ast}/r_{\text{g}} -1}$ near the horizon, $-r^{\ast} \gg r_{\text{g}}$, one obtains an analytic asymptotic solution,
\begin{equation} \displaystyle
\psi_s \simeq \sum_{n=0}^3 C_n \left( \frac{r^{\ast}}{r_{\text{g}}} \right)^n
+ \sum_{i=1}^3 f_i (r^{\ast}, C_n) \epsilon^i + {\mathcal O} (\epsilon^{4}),
\label{010801}
\end{equation}
where $f_i$ ($i=1,2,3$) are fourth-order polynomials in $r^{\ast}/r_{\text{g}}$ and an expansion coefficient, $\epsilon$, is a function of $r^{\ast}$,
\begin{equation}
\epsilon (r^{\ast}) \equiv e^{(r^{\ast}-r_{\text{g}})/2r_{\text{g}}} \frac{r_{\text{g}}}{\ell_{\text{Pl}}} \equiv e^{(r^{\ast} - r^{\ast}_{\text{Pl}})/2 r_{\text{g}}},
\end{equation}
where the sequential solutions (\ref{010801}) are good approximations for $|\epsilon (r^{\ast})| < 1$ ($r^{\ast} <
r^{\ast}_{\text{Pl}} \equiv -r_{\text{g}} \ln{\left[ r_{\text{g}}^2/ (e \ell_{\text{Pl}}^2) \right]}$).
\begin{figure}[t]
  \begin{center}
\includegraphics[keepaspectratio=true,height=46mm]{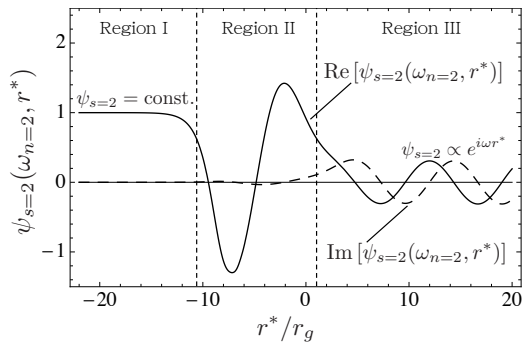}
  \end{center}
\caption{
A mode function with $C = 1, \gamma = 0$, $r_{\text{g}}/\ell_{\text{Pl}} = 10^{2.5}$, and
its frequency is the third QNM, $\omega = \omega_{n=2}$, the value of which is
$r_{\text{g}} \omega_{n=2} = 0.6705-i 10^{-2.48}$.
}%
  \label{fig2}
\end{figure}
The explicit forms of $f_i$ ($i=1,2,3$), which has the dependence on the parameters $C$ and $\gamma$,
are shown in Appendix \ref{app:b}. As is shown in (\ref{010801}), in the vicinity of the horizon
($r^{\ast} \lesssim r^{\ast}_{\text{Pl}}$), the solution of the modified RW equation is no longer oscillatory.
Hence, we have to appropriately impose a substitute boundary condition
to obtain the QNMs of a black hole with the assumed modified DRs.
Since GWs cannot propagate inward anymore at the position where $K = K_{\text{max}}$ and
the group velocity of GWs becomes zero, we will impose a condition that there is no flux for
$r^{\ast} \to - \infty$, i.e. all the ingoing energy is either reflected or absorbed by the horizon microstructure.
This boundary condition is also motivated by the firewall \cite{Almheiri:2012rt} and fuzzball \cite{Lunin:2001jy,Lunin:2002qf} picture in which there is no black hole interior.
The no flux condition at the horizon means that the mode function should be spatially constant for $r^{\ast} \to - \infty$,
and on the other hand, we can impose the outgoing condition for $r^{\ast} \to + \infty$:
\begin{equation}
\displaystyle \lim_{r^{\ast} \to - \infty} \psi_s = \text{const.}, \ \displaystyle \lim_{r^{\ast} \to + \infty}
\psi_s \sim e^{i \omega r^{\ast}}.
\label{180119}
\end{equation}
The no flux condition at the horizon singles out the solution with $C_0 \neq 0$ and $C_1=C_2 =C_3 =0$
in (\ref{010801}) since the spatial derivative of the mode function exponentially approaches zero in the limit of $r^{\ast} \to - \infty$.
Setting $\ell = s = 2$, the numerical solution for $\psi_{s = 2} (\omega, r^{\ast})$ is shown in Fig. \ref{fig2}.
In the region between the horizon ($r^{\ast} = - \infty$) and $r^{\ast} \sim r^{\ast}_{\text{Pl}}$ (Region I in Fig. \ref{fig2}),
the mode function is almost spatially constant, and there is vanishing flux.
On the other hand, in Region II in Fig. \ref{fig2}, a long-lived mode, which is the superposition of an outgoing and ingoing mode, is trapped.
Denoting the amplitudes of outgoing and ingoing modes in Region II as $A_{\text{out}}$ and $A_{\text{in}}$, respectively,
the several values of reflection rate, $|A_{\text{out}}|/|A_{\text{in}}|$, are plotted with the various values of frequency in Fig. (\ref{fig6}). Finally, Region III is located outside the angular momentum barrier, where we have imposed a purely outgoing condition. 

\section{QNMs, ringdown GWs, and reflection near the horizon}
\label{sec:QNMs}
\begin{figure}[t]
  \begin{center}
\includegraphics[keepaspectratio=true,height=45mm]{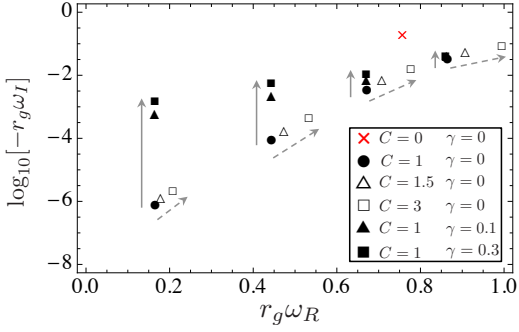}
  \end{center}
\caption{
The first four QNMs ($n= 0,1,2,3$) for $C= 1, 1.5, 3$ and for $\gamma =0, 0.1,0.3$
and the lowest-lying QNM for $C=0, \gamma = 0$ are shown. One can observe
the long-lived modes of which the imaginary parts are much smaller than unity,
$-6 \lesssim \log_{10}{|r_{\text{g}} \omega_I |} \lesssim -2$.
  }%
  \label{fig3}
\end{figure}

Numerically solving the modified RW equation (\ref{010201}),
with the boundary condition (\ref{180119}), we obtain the QNMs
that include highly long-lived modes (see Fig. \ref{fig3}). For simplicity, we here use
the P$\ddot{\text{o}}$schl-Teller (PT) potential, $V_{\text{PT}} (r^{\ast})$, which has the form \cite{Ferrari:1984zz,Poschl:1933zz}
\begin{equation}
V_{\text{PT}} (r^{\ast}) = \frac{V_0}{\cosh^2\left[\alpha (r^{\ast}-r^{\ast}_{\text{top}})/r_{\text{g}})\right]},
\end{equation}
where $V_0$, $\alpha$ and $r^{\ast}_{\text{top}}$ are constant parameters.
To mimic the angular momentum barrier for a Schwarzschild black hole $V_{2,2}(r^{\ast})$, we set $V_0 = V_{2,2}(r^{\ast}_{\text{top}})$ and
$\alpha^2 = -(2V_{2,2}(r^{\ast}_{\text{top}}))^{-1}
\partial_{r^{\ast}}^2 V_{2,2}(r^{\ast})|_{r^{\ast} = r^{\ast}_{\text{top}}}$,
where $r^{\ast}_{\text{top}}$ should be taken so that $d V_{2,2}(r^{\ast}_{\text{top}})/dr^{\ast}=0$.
In the following, we therefore choose $V_0 = 0.605/r_{\text{g}}^2$, $\alpha=0.362$ and $r^{\ast}_{\text{top}} = 1.195 r_{\text{g}}$ to mimic the RW potential. We further assume $r_{\text{g}}/\ell_{\text{Pl}} = 10^{2.5}$ for illustrative purposes. Using more realistic values of $r_{\text{g}}/\ell_{\text{Pl}} \sim 10^{40}$ only logarithmically increases echo time delays, $\Delta t_{\text{echo}}
\simeq 2 \times |r^{\ast}_{\text{Pl}}| \propto \log{\left[ r_{\text{g}} / \ell_{\text{Pl}} \right]}$
(see also \cite{Conklin:2017lwb,Wang:2018gin}).

One can see that the low-lying QNMs become less long lived as the dissipation factor $\gamma$
becomes larger, (solid arrows in Fig. \ref{fig3}). The real parts of low-lying QNMs become larger,
while their imaginary parts also become slightly larger as the parameter $C$ increases
(dashed arrows in Fig. \ref{fig3}).
\begin{figure*}[t]
  \begin{center}
\includegraphics[keepaspectratio=true,height=45mm]{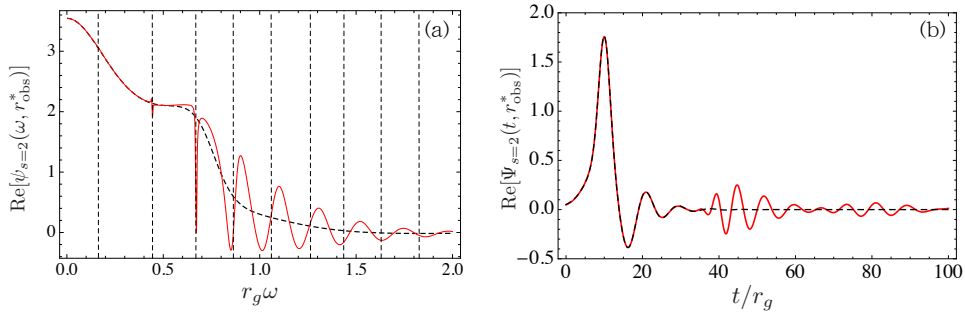}
  \end{center}
\caption{
(a): The plots of $\text{Re}[\psi_{s=2} (\omega, r^{\ast}_{\text{obs}})]$ for $C=1, \gamma=0$ (red line)
and for $C=0, \gamma=0$ (black dashed line) are shown.
The dotted line indicates the real parts of QNMs with $n=0,1,2,...,8$.
(b): The time-domain functions, $\text{Re}[\Psi_{s=2} (t, r^{\ast}_{\text{obs}})]$,
for $C=1, \gamma=0$ (red line) and for $C=0, \gamma=0$ (black dashed line) are shown.
Here we set the parameters characterizing the initial data of GWs as $(r^{\ast}_{\text{ini}}/r_{\text{g}}, \sigma/r_{\text{g}}, r^{\ast}_{\text{obs}}/r_{\text{g}}) = (3, 2, 25)$.
}
\label{fig4}
\end{figure*}

\begin{figure}[b]
  \begin{center}
\includegraphics[keepaspectratio=true,height=48mm]{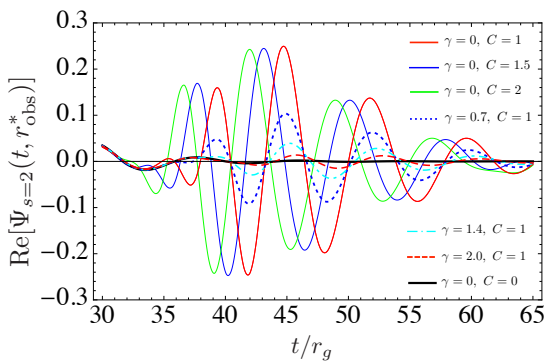}
  \end{center}
\caption{
The dependence of ringdown GWs on the parameter $C$ and on the dissipation factor $\gamma$.
Here we set the parameters for the initial data of GWs as $(r^{\ast}_{\text{ini}}/r_{\text{g}}, \sigma/r_{\text{g}}, r^{\ast}_{\text{obs}}/r_{\text{g}}) = (3, 2, 25)$.
}
  \label{fig5}
\end{figure}

We can confirm that the long-lived QNMs lead to the appearance of echoes in the late-time tail of
ringdown GWs by calculating the time-domain wave function, which can be recovered via
\begin{equation}
\Psi_{s=2} (t,r^{\ast}) = \frac{1}{\sqrt{2 \pi}} \int d \omega e^{-i \omega t} \psi_s (\omega, r^{\ast}).
\end{equation}
Assuming a static initial data, $\partial_t \Psi_s (0, r^{\ast}) = 0$, the mode function $\psi_s$ can be obtained by the (retarded) Green's function, $G(r^{\ast}, r^{\ast} {}', t)$, as \cite{Berti:2009kk}
\begin{equation}
\begin{split}
\psi_{s=2} (\omega, r^{\ast}) &= \int dr^{\ast} {}' \partial_t G (r^{\ast}, r^{\ast} {}', t) \Psi_{s=2} (0,r^{\ast} {}')\\
&=\frac{\psi_+}{2i \omega A_{\text{in}}} \int^{r^{\ast}}_{- \infty} dr^{\ast} {}' I (\omega, r^{\ast} {}') \psi_- \\
&~~~~~+ \frac{\psi_-}{2i \omega A_{\text{in}}} \int^{\infty}_{r^{\ast}} dr^{\ast} {}' I (\omega, r^{\ast} {}') \psi_+,
\end{split}
\end{equation}
where $I (\omega , r^{\ast})$ is a source term that includes initial data of GWs around
the black hole.
Here, we take the static and Gaussian initial data, which have a peak at $r^{\ast} = r^{\ast}_{\text{ini}}$ with
its dispersion $\sigma$, that is,
$\Psi_s (t=0, r^{\ast}) = e^{-(r^{\ast} - r^{\ast}_{\text{ini}})^2 / \sigma^2}$ and
$\partial_t \Psi_s (t =0, r^{\ast}) = 0$, and the form of the source term is given by
\begin{equation}
I (\omega , r^{\ast}) = i \omega \Psi_{s=2} (t=0, r^{\ast}).
\end{equation}
The real part of the mode function and time-domain wave function, as seen by an observer at $r^{\ast} = r^{\ast}_{\text{obs}}$, $\text{Re}[\psi_{s=2} (\omega, r^{\ast}_{\text{obs}})]$ and
$\text{Re}[\Psi_{s=2} (t, r^{\ast}_{\text{obs}})]$, are shown in Fig. (\ref{fig4}).
One can find that the long-lived QNMs with $n=3,4,5$ are relatively excited compared to
other QNMs (red line in Fig. \ref{fig4}a), which leads to the echoes in the late-time
tail of ringdown GWs.
We find out that the phase of echoes depends on the parameter $C$ (Fig. \ref{fig5}) and
the time interval in which we observe the echo, $\Delta t_{\text{echo}}$,
can be evaluated as (see Appendix \ref{app:c})
\begin{equation}
\Delta t_{\text{echo}} \simeq -2 \times r^{\ast}|_{K\simeq K_{\rm max}} \simeq
2 r_{\text{g}} \ln{\left[ 2 (2C^2+\gamma^2) E_{\text{Pl}}^2 \over e (4C^2+\gamma^2)^2 \omega^2 \right]}. 
\label{techo}
\end{equation}
On the other hand, the amplitude of echoes depends on the dissipation factor $\gamma$,
and the echoes disappear in the limit of $|\gamma| \gg |C|$ (Fig. \ref{fig5}).
This is consistent with the dependence of QNMs on the parameters $C$ and $\gamma$;
the real parts of QNMs (the phase of echoes) depend on $C$ and the imaginary parts
(the dissipation effect on ringdown GWs) largely depend on $\gamma$.

We also calculated the reflection rate of GWs near the black hole horizon for several frequencies of GWs (Fig. \ref{fig6})
and confirmed that the reflection rate is consistent with the amplitude of echo (see Appendix \ref{app:d}).
A mode function $\psi_{s=2} (\omega,r^{\ast})$ in the range of $r^{\ast}_{\text{Pl}} < r^{\ast} < 0$
(Region II in Fig. \ref{fig2}) can be decomposed into
an outgoing and ingoing mode: $\psi_{s=2} = A_{\text{out}} e^{i \omega r^{\ast}} + A_{\text{in}} e^{-i \omega r^{\ast}}$.
Then, we can calculate the reflection rate, $|A_{\text{out}}| / |A_{\text{in}}|$, as a function of $\omega$,
and one finds that the GWs is perfectly reflected at the horizon for $\gamma = 0$,
while the reflection rate drops as the dissipation term, $\gamma$, increases (Fig. \ref{fig6}a).
Interestingly, the reflection rate of long-wavelength GWs are higher compared to those of short-wavelength GWs,
which is consistent with Ref. \cite{Saravani:2012is}. We can find an
analytic form of the reflection rate based on the WKB approximation (see Appendix \ref{app:c}),
\begin{equation}
|A_{\text{out}}| / |A_{\text{in}}| \simeq \exp{\left[ -   \frac{\sqrt{2 + 4 C^2/\gamma^2}}{2\pi(1 + 4 C^2/\gamma^2)}\left(\hbar \omega \over k_B T_H\right)  \right]},
\label{wkb_hawking}
\end{equation}
in terms of Hawking temperature $T_H= \hbar\kappa/(2\pi k_B)$ or surface gravity $\kappa$ for the black hole horizon.  

The zero flux boundary condition (\ref{180119}) is not appropriate for a modified DR with $\gamma =0$ and with $C^2 < 0$ since
it does not lead to zero group velocity of infalling GWs near horizon.
However, the wave number at which the group velocity becomes zero is given by $K_{\text{max}} = E_{\text{Pl}} / \sqrt{\gamma^2/2 + 2 C^2}$,
and therefore, the negativity of $C^2$, as suggested by e.g., HL gravity \cite{Vacaru:2010rd},
can be offset by the large value of $\gamma$ for which
the zero flux condition is satisfied. Fixing $\gamma$ with a large value, we here investigate the dependence of the reflection rate on the parameter
$C$  (Fig. \ref{fig6}b). Since the term including the parameter $C$ does not contribute to the reflection of GWs in the case of $C^2 < 0$, the reflection rate
is smaller compared to the case of $C^2 > 0$ or $C=0$. Furthermore, taking $C^2 < 0$ and $\gamma^2 \gg |C^2|$
suppresses the amplitude of echoes and shifts them to lower frequencies. 
\begin{figure}[b]
  \begin{center}
\includegraphics[keepaspectratio=true,height=55mm]{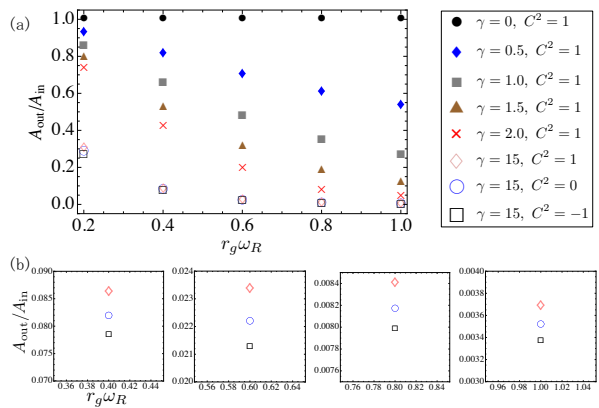}
  \end{center}
\caption{
The several values of reflection rate with the various values of the parameters and with the various values of frequency, $\omega_{R}$,
with $\omega_{I} = 0$.
}
  \label{fig6}
\end{figure}

\section{discussions} \label{sec:discussions}
Let us now see how currently measured echo properties constrain the modified dispersion relation (\ref{18020401}).  For the binary NS merger event GW170817 \cite{TheLIGOScientific:2017qsa} , Ref. \cite{Abedi:2018npz} reports repeating echoes with frequency $f= 72$ Hz that decay (in power) by a factor of 2 within 0.2 sec. This result is consistent with a black hole remnant with $M=2.6-2.7 M_{\odot}$ and spin $0.84-0.87$, fixing the exponent in (\ref{wkb_hawking}) to -2.4\% for $\frac{\hbar \omega}{2\pi k_B T_H} \simeq 0.035$, which in turn implies $\gamma \simeq C$.  

Furthermore, Ref. \cite{Conklin:2017lwb} reports echoes in aLIGO BH mergers are consistent with $2\Delta t_{\rm echo}/r_{\text{g}} = 620 \pm 32$. For the aLIGO frequency range $100-200$ Hz, Eq. (\ref{techo}) implies $K_{\rm max} \sim  E_{\text{Pl}}/ C \sim 10^{-6 \pm 2} E_{\text{Pl}} = 10^{13 \pm 2}$ GeV, which is suggestively close to both the scale of grand unified theories, as well as the tachyacoustic big bang models that seed cosmic structures \cite{Agarwal:2014ona,Afshordi:2016guo}.

Blas \textit{et al.} \cite{Blas:2010hb} combined current bounds on Einstein-Aether/HL theories with technical naturalness arguments to suggest an upper limit on the Lorentz breaking scale, $E_{\text{Pl}} / C < 10^{16}$ GeV. On the other hand, direct gravitational wave observations give lower a bound $E_{\text{Pl}} / C > 10 $ meV (e.g., \cite{TheLIGOScientific:2016src}).
Our present measurement of $10^{13 \pm 2}$ GeV is consistent with both these constraints.
Ref. \cite{Nielsen:2018lkf} reports, however, that, although there is some evidence echoes in LVT151012 and GW151226, as both have positive log Bayes factors (Table II in \cite{Nielsen:2018lkf}), this evidence is not very strong since the magnitude of the factors is comparable to or smaller than unity. Therefore, note that the above discussion is not based on conclusive evidence for the echo signals. It is expected that improvements in observational sensitivity and echo waveform models may reveal if echo signals are emitted at the final stage of merger events leading to the formation of black holes (see, e.g., \cite{Abedi:2016hgu,Ashton:2016xff,Abedi:2017isz,Conklin:2017lwb,Westerweck:2017hus,Abedi:2018npz,Abedi:2018pst} for the relevant discussion).

\section{Conclusions} \label{sec:conclusions}
In this manuscript, we have proposed the possibility that the microstructure on black hole horizons
causes a Planck-scale modification of DR,
which could be probed by observing the late-time tail of ringdown GWs from the formation of black holes.
Assuming a simplified modified DR with the characteristic dissipation and dispersion scales,
$\gamma$ and $C$,
we have numerically calculated the QNMs of a Schwarzschild black hole.
The modified DRs change the boundary condition for mode functions of GWs near the horizon
and  lead to the highly long-lived QNMs, resulting in the
appearance of echoes at the late-time tail of ringdown GWs.
Although the echoes in ringdown GWs have been studied in the context of the probe of exotic compact objects \cite{Visser:1995cc,Mazur:2001fv,Mathur:2005zp,Almheiri:2012rt,Holdom:2016nek,Saravani:2012is}, our results suggest that the echoes could be a probe of quantum gravity theory itself, which would pinpoint the modification of DR in nature.
In particular, current reported echo signals are consistent with a Lorentz-violation scale of $10^{13 \pm 2}$ GeV, with comparable dissipation and dispersion.
The future work will explore the ringdown GWs with the DRs predicted by the loop quantum gravity \cite{Girelli:2012ju} and the HL gravity (including the dynamics of the khronon field) \cite{Vacaru:2010rd}.

\begin{acknowledgments}
N. O. is supported by Grant-in-Aid for JSPS Fellow Grant No. 16J01780. N. A. is supported by the Perimeter Institute and Natural Sciences and Engineering Research Council of Canada (NSERC). Research at the Perimeter Institute is supported by the Government of Canada through the Department of Innovation, Science and Economic Development Canada, and
by the Province of Ontario through the Ministry of Research
and Innovation.
\end{acknowledgments}

\appendix

\section{Derivation of $K_{\text{max}} = E_{\text{Pl}}/\sqrt{2 C^2 + \gamma^2/2}$}
\label{app:a}
Starting with the modified DR,
\begin{equation}
\Omega^2 = K^2 -C^2 K^4/E_{\text{Pl}}^2 - i \gamma K^2 \Omega / E_{\text{Pl}},
\label{062701}
\end{equation}
one has 
\begin{equation}
\Omega = \pm K \sqrt{1-(C^2 + \gamma^2/4) K^2/ E_{\text{Pl}}^2} -i \frac{\gamma K^2}{2 E_{\text{Pl}}}. \label{freq}
\end{equation}
Calculating $d \Omega/ dK$, one obtains
\begin{equation}
\frac{d \Omega}{d K} = \pm \frac{1- (2 C^2 + \gamma^2/2) K^2 / E_{\text{Pl}}^2}
{\sqrt{1-(C^2 + \gamma^2/4) K^2/ E_{\text{Pl}}^2}} - i \gamma \frac{K}{E_{\text{Pl}}},
\end{equation}
where the first term, $\text{Re} (d \Omega/d K)$, is the group velocity and the second term, $\text{Im} (d \Omega/d K)$,
gives the dissipation rate.
Finally, one finds that $K_{\text{max}}$, for which the group velocity becomes zero, takes the form of
\begin{equation}
K_{\text{max}} = \frac{E_{\text{Pl}}}{\sqrt{2 C^2 + \gamma^2 /2}}.
\end{equation}

\section{Asymptotic solutions for the modified RW equation}
\label{app:b}
In the near-horizon limit, $r^{\ast} \to - \infty$, the analytic asymptotic solutions for the modified RW equation can be derived.
Since the term including a fourth derivative in Eq. (4) becomes the most dominant term in the near-horizon limit,
the zeroth order of the asymptotic solutions should be a fourth-order polynomial.
Furthermore, it is sequentially found that the first and second orders of the asymptotic solutions
should be suppressed by $e^{r^{\ast}/2r_g}$ and by $e^{r^{\ast}/r_g}$, respectively,
\begin{equation}\displaystyle
\psi_s (\omega, r^{\ast}) = \sum_{i=0}^{3} f_i(r^{\ast}) \epsilon^i + \mathcal{O} (\epsilon^4),
\label{062704}
\end{equation}
where $\epsilon (r^{\ast}) \equiv r_{\text{g}} e^{(r^{\ast} - r_{\text{g}})/2r_{\text{g}}} /\ell_{\text{Pl}}$ and $f_i (r^{\ast})$ ($i = 0,1,2,3$) are fourth-order polynomials.
Substituting the following ansatz into the modified RW equation, Eq. (4),
\begin{align}
f_0 (r^{\ast}) &\equiv \displaystyle \sum_{n=0}^{3} C_n (r^{\ast}/r_g)^n,\\
f_1 (r^{\ast}) &\equiv  \sum_{n=0}^{3} D_n (r^{\ast}/r_g)^n,\\
f_2 (r^{\ast}) &\equiv  \sum_{n=0}^{3} E_n (r^{\ast}/r_g)^n,\\
f_3 (r^{\ast}) &\equiv  \sum_{n=0}^{3} F_n (r^{\ast}/r_g)^n,
\end{align}
one can obtain all coefficients of the ansatz, $\{ D_n \}_{n=0,1,2,3}$, $\{ E_n \}_{n=0,1,2,3}$, and $\{ F_n \}_{n=0,1,2,3}$ as follows:
\begin{widetext}
\begin{align}
D_0 &= \frac{32 i \gamma \bar{\omega}}{C^2} (C_2-24 C_3),\\
D_1 &= \frac{96iC_3 \gamma \bar{\omega}}{C^2},\\
D_2 &=D_3=0,\\
E_0 &= \frac{1}{C^4} \left( -8 (C_2-24 C_3) \gamma^2 \bar{\omega}^2 +C^2 \left( -(C_0 -4 C_1) \bar{\omega}^2
+24 C_3 (1+5 \bar{\omega}^2) -2 C_2 (1+10 \bar{\omega}^2) \right) \right),\\
E_1 &= \frac{-1}{C^4} \left( 24 C_3 \gamma^2 \bar{\omega}^2 +C^2 \left( (C_1 - 8 C_2) \bar{\omega}^2 +6 C_3
(1+10 \bar{\omega}^2) \right) \right),\\
E_2 &= - \frac{\bar{\omega}^2}{C^2} (C_2 - 12 C_3),\\
E_3 &= - \frac{C_3 \bar{\omega}^2}{C^2},\\
F_0 &= - \frac{32 (C_2 - 8 C_3) e^{1/2}}{81 C^2 r_g} - \frac{16 i \gamma \bar{\omega} \left( 10 C_2 -188 C_3 + C_0 \bar{\omega}^2 \right)}{81 C^4}
+ \frac{32 i \left( 21 C_1 -259 C_2 + 5332 C_3 \right) \gamma \bar{\omega}^3}{729 C^4}\nonumber \\
&- \frac{128 i \left( C_2 -26 C_3 \right) \gamma^3 \bar{\omega}^3}{81 C^6},
\end{align}
\end{widetext}
\begin{widetext}
\begin{align}
F_1 &= - \frac{32 C_3 e^{1/2}}{27 C^2 r_g} - \frac{160 i C_3 \gamma \bar{\omega}}{27 C^4}- \frac{16 i \left( 3 C_1 -28 C_2 +518 C_3 \right)
\gamma \bar{\omega}^3}{243 C^4} - \frac{128 i C_3 \gamma^3 \bar{\omega}^3}{27 C^6},\\
F_2 &=- \frac{16 i \left( C_2 - 14 C_3 \right) \gamma \bar{\omega}^3}{81 C^4},\\
F_3 &=- \frac{ 16 i C_3 \gamma \bar{\omega}^3}{81 C^4},
\end{align}
\end{widetext}
where $\bar{\omega} \equiv r_{\text{g}} \omega$ and $\{ C_n \}_{n=0,1,2,3}$ are arbitrary constants to be determined by a boundary condition for the
modified RW equation. Here we impose $C_0 \neq 0$ and $C_1=C_2=C_3=0$ as the no-flux condition near the horizon
($r^{\ast} \ll r^{\ast}_{\text{Pl}}$).
Finally, one has the asymptotic form of the mode function:
\begin{equation}
\psi_{s=2} (\omega, r^{\ast}) \simeq C_0 \left( 1 - \frac{\bar{\omega}^2}{C^2} \epsilon^2
-i \frac{16}{81} \gamma \frac{\bar{\omega}^3}{C^4} \epsilon^3 \right) \ \ \text{for} \ r^{\ast} < r^{\ast}_{\text{Pl}}.
\label{mode_func_approx1}
\end{equation}
One may find that the imaginary part of the asymptotic form of the mode function is proportional to the dissipation factor,
$\text{Im} (\psi_{s=2}) \propto \gamma$, which suggests that the reflection rate may decrease as $\gamma$ increases
as is shown in the next section.

\section{Reflection rate of GWs near the horizon}
\label{app:c}
\begin{figure}[t]
  \begin{center}
\includegraphics[keepaspectratio=true,height=42mm]{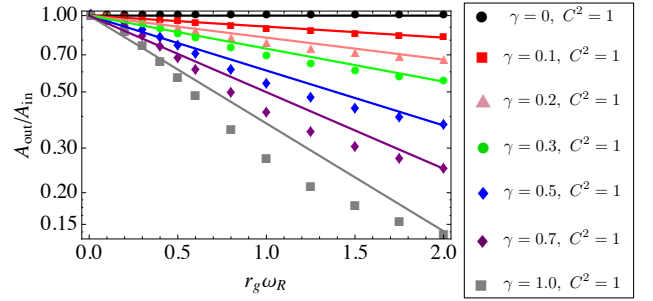}
  \end{center}
\caption{
The comparison between the semianalytic form ( \ref{wkb}) and the numerical calculated reflection rate near the horizon.
}
  \label{app_1}
\end{figure}
In this Appendix, we will derive the analytic form of the reflection rate of GWs near the horizon,
$|A_{\text{out}}| / |A_{\text{in}}|$, by using the WKB approximation.

To see this, we note that the imaginary part of the frequency in Eq. (\ref{freq}) denotes the rate of decay in the amplitude of a wave packet. As such, in the WKB limit, we can write:
\begin{eqnarray}
\ln\left(\frac{|A_{\text{out}}|}{|A_{\text{in}}|}\right) \simeq - \oint dt \sqrt{F(r^\ast)} \frac{\gamma K^2}{2 E_{\text{Pl}}} \nonumber\\ \simeq - \frac{\gamma \omega^2}{2 E_{\text{Pl}}} \oint \frac{|dr^\ast|}{\sqrt{F(r^\ast)}} \simeq -\frac{2 \gamma r_{\text{g}} \omega^2}{E_{\text{Pl}} \sqrt{F(r^\ast_{\rm min})}}, \label{WKB}
\end{eqnarray}
where $\oint$ is the integral over the classical wave packet trajectory with $\Omega(r)=\omega/\sqrt{F(r)}$, and we {\it only} use the change in the dispersion relation to compute the reflection radius \footnote{Given that $\Omega(K)$  in Eq. (\ref{freq}) is complex, $r^\ast_{\rm min}$ resulting from $\Omega(K_{\rm max})=\omega/\sqrt{F(r^\ast_{\rm min})}$ is generally not real. As such, here we use
$|\Omega|(K_{\rm max}) = \omega/\sqrt{F(r^\ast_{\rm min})}$ to define a real reflection radius, $r^\ast_{\rm min}$. }:
\begin{equation}
r^\ast_{\rm min} \equiv r^{\ast}|_{K\simeq K_{\rm max}}  \simeq -r_{\text{g}} \ln{\left[ 2 (2C^2+\gamma^2) E_{\text{Pl}}^2 \over e (4C^2+\gamma^2)^2 \omega^2 \right]}.
\end{equation}
Plugging this into Eq. (\ref{WKB}) yields
\begin{equation}
|A_{\text{out}}| / |A_{\text{in}}| \simeq \exp{\left[ -2   r_{\text{g}} \omega \frac{\sqrt{2 + 4 C^2/\gamma^2}}{1+ 4 C^2/\gamma^2} \right]}.
\label{wkb}
\end{equation}
\\
The analytic formula (\ref{wkb}) and the numerically obtained reflection rates are compared in Fig. \ref{app_1}, which appear to agree for small values of $\gamma \omega r_{\text{g}}$, where the adiabatic condition is satisfied.

We can also relate $r_{\text{g}}$ to surface gravity $\kappa$, or Hawking temperature of $T_H$ via $k_B T_H \equiv \hbar \kappa/(2\pi) = \hbar/(4\pi r_{\text{g}})$. Therefore, Eq. (\ref{wkb}) can be written in terms of Hawking temperature as
\begin{equation}
|A_{\text{out}}| / |A_{\text{in}}| \simeq \exp{\left[ -   \frac{\sqrt{2 + 4 C^2/\gamma^2}}{2\pi(1 + 4 C^2/\gamma^2)}\left(\hbar \omega \over k_B T_H\right)  \right]}.
\label{wkb_hawking}
\end{equation}  
Given that the reflection only depends on the local blueshift near the horizon, we expect this formula to be valid for spinning BHs as well. 
\begin{figure}[b]
  \begin{center}
\includegraphics[keepaspectratio=true,height=45mm]{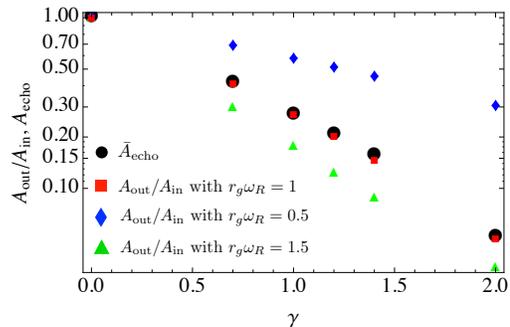}
  \end{center}
\caption{
The comparison between the regularized amplitude of the first echo
and the reflection rate with $r_g \omega_R = 0.5$, $1$, and $1.5$
(blue, red, and green points, respectively).
}%
  \label{app_2}
\end{figure}

\section{Consistency between the reflection rate and amplitude of GW echoes}
\label{app:d}
As supporting evidence that the GW echoes originate from the reflection of GWs at the horizon, we confirmed that the dependence of the amplitude of the first echo on the dissipation effect $\gamma$ (Fig. \ref{app_2}). As is shown in Fig. 3, the most excited QNM is around $r_g \omega_R \sim 1$, and this mode may be dominant in the GW echoes.
This is consistent with the comparison with the reflection rate for $r_{\text{g}} \omega_R =1$ (red points)
shown in Fig. \ref{app_2}, in which one may find that the reflection rate with $r_g \omega_R = 1$ fits the regularized amplitude of the first echo $\bar{A}_{\text{echo}}$ well.
On the other hand, the reflection rate for the modes of $r_{\text{g}} \omega_R = 0.5$ (blue points) and $1.5$ (green points),
which are not excited well in the GW echoes, do not fit the regularized amplitude of the echo.

\end{document}